\documentclass[journal,twoside,web]{ieeecolor}
\usepackage{generic}
\usepackage{cite}
\usepackage{amsmath,amssymb,amsfonts}
\usepackage{algorithmic}
\usepackage{graphicx}
\usepackage{textcomp}
\def\BibTeX{{\rm B\kern-.05em{\sc i\kern-.025em b}\kern-.08em
    T\kern-.1667em\lower.7ex\hbox{E}\kern-.125emX}}

\begin{document}

\title{EEG-Based Analysis of Brain Responses in Multi-Modal Human-Robot Interaction: Modulating Engagement}

\author{Suzanne Oliver, \textit{Graduate Student Member, IEEE}, Tomoko Kitago, Adam Buchwald,  S. Farokh Atashzar$^{\dag}$,  \textit{Senior Member, IEEE}
\thanks{Suzanne Oliver is with the Department of Mechanical and Aerospace Engineering, New  York University (NYU), New York, NY. Tomoko Kitago is with the Department of Neurology, New York Medical College, Valhalla, NY, and the Department of Neurology, Westchester Medical Center, Valhalla, NY. Adam Buchwald is with the Department of Communicative Sciences and Disorders, NYU. S. Farokh Atashzar is with the  Departments of Electrical and Computer Engineering, Mechanical and Aerospace  Engineering, and Biomedical Engineering, NYU.  Atashzar is also with NYU WIRELESS Center and NYU CUSP.}
\thanks{\dag  Corresponding author: Atashzar (f.atashzar@nyu.edu).}%
\thanks{This material is partially supported by the U.S. National Science Foundation (NSF) under Awards: 2229697, 2208189. It is also supported by the U.S. National Institutes of Health (NIH) Grant R21DC019955.}
}

\maketitle

\begin{abstract}
User engagement, cognitive participation, and motivation during task execution in physical human-robot interaction are crucial for motor learning. These factors are especially important in contexts like robotic rehabilitation, where neuroplasticity is targeted. However, traditional robotic rehabilitation systems often face challenges in maintaining user engagement, leading to unpredictable therapeutic outcomes. To address this issue, various techniques, such as assist-as-needed controllers, have been developed to prevent user slacking and encourage active participation. In this paper, we introduce a new direction through a novel multi-modal robotic interaction designed to enhance user engagement by synergistically integrating visual, motor, cognitive, and auditory (speech recognition) tasks into a single, comprehensive activity. To assess engagement quantitatively, we compared multiple electroencephalography (EEG) biomarkers between this multi-modal protocol and a traditional motor-only protocol. 
Fifteen healthy adult participants completed 100 trials of each task type. Our findings revealed that EEG biomarkers, particularly relative alpha power, showed statistically significant improvements in engagement during the multi-modal task compared to the motor-only task. Moreover, while engagement decreased over time in the motor-only task, the multi-modal protocol maintained consistent engagement, suggesting that users could remain engaged for longer therapy sessions. Our observations on neural responses during interaction indicate that the proposed multi-modal approach can effectively enhance user engagement, which is critical for improving outcomes. This is the first time that objective neural response highlights the benefit of a comprehensive robotic intervention combining motor, cognitive, and auditory functions in healthy subjects.
\end{abstract}


\section{Introduction}
\IEEEPARstart{H}{uge} numbers of Americans are affected by neurological diseases that necessitate physical rehabilitation. In particular, stroke is a leading cause of disability in America and can cause issues with motor control, memory, and cognition. Each year, approximately 800,000 Americans experience a stroke, and that number is projected to continue to rise throughout the next decade \cite{martin20242024, pu2023projected}. As a result, the development of new and improved rehabilitation devices is critical. 

One rapidly growing area in this field is robotic platforms for motor learning \cite{maciejasz2014survey, heuer2015robot, asin2020haptic, brown2016designing, cherry2017expanding, akbari2021robotic}. Motor learning is a proposed mechanism underlying training-related recovery to help people with neurological impairments regain their motor control \cite{kitago2013motor}.
However, despite the potential benefits of robotic rehabilitation, the evidence supporting the effectiveness of robotic motor learning is mixed \cite{rodgers2019robot, de2024upper}. One key reason for this is known to be unpredictable cognitive engagement and participation of patients in task performance. Phenomena such as slacking can potentially reduce the stimulation of neural pathways and potential benefits regarding motor learning in general \cite{blank2014current, kaelin2005role}. On the other hand, increased engagement corresponds to higher levels of neuroplasticity, meaning the brain is more susceptible to reorganization and adaptation, which are essential for learning and retaining motor skills \cite{danzl2012facilitating,blank2014current, krupinski2014towards, kayes2022developing, levin2015emergence}. Increasing the user's attention to task conduction and providing motivation are key factors in increasing engagement and, thus, neuroplasticity.

Moreover, engagement also affects patient's adherence to motor therapy programs. Lack of engagement is linked to poor adherence to physical therapy regimens, meaning patients are more likely to skip sessions \cite{teo2022identifying, lohse2013video}. Since motor learning depends on consistent repetitions to regain motor skills, this low adherence reduces the effectiveness. Thus, engagement is critical not just to maximize the potential benefit of each therapy session through increased neuroplasticity but also simply to encourage patients to complete their therapy sessions.

With this in mind, there have been many attempts to improve engagement in robotic motor learning platforms. For instance, using an assist-as-needed protocol, where the robot only helps the user complete the desired movement if they are unable to do it on their own, has been shown to increase engagement levels by providing subjects with a greater sense of agency \cite{li2021engagement, shahbazi2016robotics, blank2014current}. These assist-as-needed protocols prevent user slacking and require users to actively participate in the task as much as possible, which results in improved encoding of motor skills to the brain's motor cortex \cite{kaelin2005role}. These robots can also adjust the difficulty of the task based on user performance to keep the user motivated, combating the issue of poor adherence to robot therapy regimes \cite{blank2014current}.
Additionally, many virtual reality platforms and video games have been designed to immerse users in the motor learning paradigm and potentially add a sense of realism, with the aim of increasing engagement \cite{lohse2013video, levin2015emergence, zimmerli2013increasing}. The motivation in both the assist-as-needed and virtual reality cases is to improve patient outcomes by increasing engagement.

Given the importance of engagement in motor learning, it is critical to create rehabilitation strategies that engage the user. For this purpose, in this paper, we proposed the use of a multi-modal robotic motor learning exercise with the goal of enhancing the engagement and cognitive participation of human subjects during task conduction. In this task, which was adapted from a study on post-stroke aphasia treatment \cite{fridriksson2018transcranial}, users had to compare visual and verbal stimuli and move a robot handle to indicate if the two stimuli matched. This was compared to a motor-only task, where the participants only had to move the robot handle to a specified location. 
We hypothesized that the added cognitive and sensory processing components in the matching task would increase user engagement compared to traditional motor-only task.

However, it can be difficult to quantify how well these platforms are engaging users. Typically, questionnaires are used to ask about user engagement, but these are subjective and can be biased. They are also limited by the time intervals when the survey is given, and asking for feedback can break the immersion and thus reduce engagement \cite{greene2015measuring, kumar2020machine}. Thus, it would be ideal to have a non-invasive, quantitative measure of engagement that is continuously recorded to avoid these issues.

For this purpose, power analysis of electroencephalography (EEG) measurements has been proposed to quantify engagement in a variety of situations. Analysis of the EEG power at different frequency bands, such as the delta band (1-4 Hertz), theta band (4-8 Hertz), and the alpha band (8-13 Hertz), have revealed characteristic changes depending on the level of subject engagement in a task.

The effect of engagement is most commonly noted in the alpha band, where an increase in engagement corresponds to a decrease in alpha band power. This effect has been seen in school settings \cite{davidesco2023detecting}, video games \cite{lim2019comparison, natalizio2024real}, virtual reality platforms \cite{magosso2019eeg}, storytelling \cite{boudewyn2018must}, and rehabilitation environments \cite{rogers2021single}. The decrease in alpha power is particularly noticeable in the occipital and parietal regions of the brain \cite{davidesco2023detecting, lim2019comparison, magosso2019eeg}. A theoretical explanation for this phenomenon is that alpha activity acts as an `inhibitor' of sensory processing, so when the brain is engaged in the task (i.e., listening, processing images, etc.), alpha decreases. In contrast, when alpha activity is higher, it may prevent distraction from external stimuli \cite{magosso2019eeg, chikhi2022eeg}.

Conversely, power in the theta and delta bands tends to increase with engagement. Theta power is associated with increased cognitive workload and can also be a marker for engagement and immersion \cite{chikhi2022eeg, lim2019comparison, rogers2021single, davidesco2023detecting, natalizio2024real}. As with the alpha band, the occipital and parietal regions of the brain are particularly sensitive to changes in engagement and attention in the theta band \cite{davidesco2023detecting, lim2019comparison, magosso2019eeg}. Delta power has also been shown to increase during more engaging tasks, such as Go/No Go tasks \cite{harmony2013functional}, and with higher emotional involvement \cite{knyazev2009event}.

Since theta and alpha power are both associated with engagement, the ratio of theta power to alpha power can also be used as a marker for engagement and cognitive attention that is sensitive to changes in both frequency bands \cite{lim2019comparison, nan2022alpha, pavithran2019index}. 

We used these EEG neural markers of engagement to investigate the engagement levels of fifteen subjects during the multi-modal matching task and the traditional motor-only task. We found that the EEG engagement biomarkers were higher for the matching task compared to the motor-only task, particularly in the moments right before the subject began to move when they were processing the visual and auditory stimuli. These results suggest the potential of using a multi-modal approach to enhance engagement in patients getting robot-assisted therapy.

\section{Methods}

\subsection{Experimental Setup}

\begin{figure*}[h]
\centering
\centerline{\includegraphics[width=0.95\textwidth]{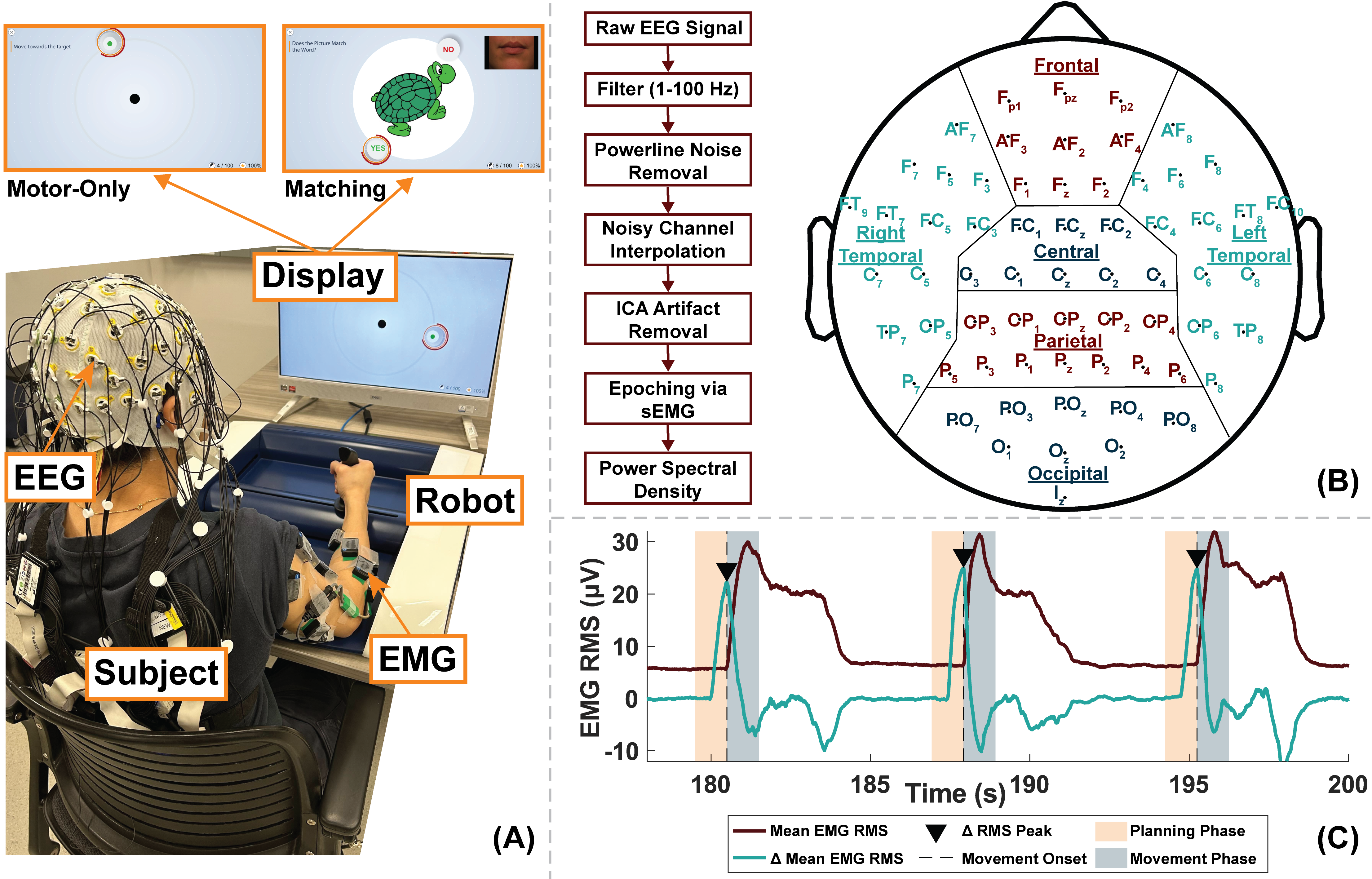}}
\caption{Overview of experiment design and processing pipeline, showing (A) the experimental setup, (B) the EEG processing steps and brain regions, and (C) an example of sEMG epoching.}
\label{experimentoverview}
\end{figure*}

Fifteen healthy adult subjects participated in this study (nine males, six females, mean age $27.5 \pm 4.2$ years). The Institutional Review Board of New York University approved the study, and all subjects signed a written consent form prior to participating in the study. All participants denied any history of musculoskeletal injury or neurological impairment.

In this study, subjects completed two tasks using a two-degree-of-freedom rehabilitation robot, the H-Man (Articares, Singapore). During task conduction, EEG and surface electromyography (sEMG) signals were recorded from the participants. The EEG data was recorded from 64 channels in the 10-10 configuration (Acticap and LiveAmp, Brain Products GmbH, Munich, Germany) at 500 Hertz. 

The sEMG readings were recorded from fifteen Bipolar Delsys Trignosystem (Delsys, Natick, MA, USA) sEMG sensors, recording at 2148 Hertz. The sEMG sensors were placed on the subject's right side, along the upper back, upper arm, and forearm. The sensors were placed on the Trapezius Descendens and Trapezius Transversalis muscles following the SENIAM guidelines \cite{frericks1999recommendations}. The sensors on the Posterior Deltoid, Lateral Deltoid, Anterior Deltoid, Long head of the Triceps, Lateral head of the Triceps, Short Head of the Biceps Brachii, Long Head of the Biceps Brachii, Palmaris Longus, Flexor Carpi Radialis, Brachioradialis, and Extensor Carpi Radialis muscles were placed following the guidelines in \cite{barbero2012atlas}. Finally, sensors were placed on the Extensor Digitorum and Extensor Carpi Ulnaris following the directions in \cite{zipp1982recommendations}.

Before beginning the experiment, subjects were seated in front of the H-man robot and a computer screen. The height of the table and position of the chair were adjusted so that the subject was centered in front of the robot, and they could reach all parts of the robot workspace without moving their torso. All subjects were right-handed and performed the task with their right arm.

Subjects were asked to complete a series of trials using the robot to control a cursor on the screen in front of them. These trials were divided into two types: Motor-Only and Matching. In the Motor-Only task, users were shown a target location on the screen, and they moved the robot handle to reach and maintain that position for two seconds. The target location changed for each trial but was always the same distance from the home position. In the Matching tasks, the users were shown an image on the screen, and a video of a word was played simultaneously. Users had to move the robot handle to select `YES' if the word matched the image or `NO' if the word did not match the image and maintain that position for two seconds. As in the Motor-Only task, the positions of the `YES' and `NO' options were randomized for each trial, but they were always the same distance from the home position and directly opposite each other. The experimental setup and both tasks are shown in Fig. \ref{experimentoverview}(A). 

For both the Motor-Only and Matching tasks, the robot was set to Resistance mode, meaning it applied a force that pulled the user back to the center home position. The force was modeled as a spring-damper system with a spring coefficient of 300 N/m and a damping coefficient of 125 N/ms. After each trial, the robot returned to the home position and waited for two seconds for the next trial to begin.
Each subject completed 100 trials of one type (either Motor-Only or Matching) in a row, took a short break, and then completed 100 trials of the other type. The order of the trial types was randomized for each participant, with eight participants completing the Motor-Only task first and the remaining seven completing the Matching task first.

\subsection{Data Processing}

\subsubsection{EEG Signal Processing}

The EEG signals were processed in MATLAB (MathWorks  Inc.,  Natick,  MA), using the EEGLAB Toolbox \cite{delorme2004eeglab}. The processing steps are shown in Fig. \ref{experimentoverview}(B). First, the raw signals were filtered using a fifth-order high-pass Butterworth filter with a cutoff of 1 Hertz, and then a sixth-order low-pass Butterworth filter with a cutoff of 100 Hertz. Line noise was removed at 60 Hertz using the EEGLAB \textit{cleanline} function with a bandwidth of 2 Hertz.

Next, channels with high noise (kurtosis $> 2$) were identified and removed. These signals were replaced by interpolating from neighboring channels via the EEGLAB \textit{interp} function. Finally, the EEGLAB \textit{run\_ica} function was used to identify and remove artifacts (such as eye blinks, heartbeat, etc.) from the signals.

\subsubsection{Epoching from sEMG}

The sEMG signals were used to define the start period of each trial. First, the signals were filtered with a fourth-order Butterworth bandpass filter between 20 and 500 Hertz. Additionally, fourth-order Butterworth bandstop filters with a width of 4 Hertz were applied at multiples of 60 Hertz to remove power line noise. Next, the Root-Mean-Square (RMS) of each signal was taken at each time step over the previous 0.5 seconds of data, which was used as a metric of the magnitude of muscle activity. The mean of these RMS values across all fifteen sensors was then taken to get the average muscle activity at each time step. 

The onset of the motion for each iteration was considered to be the time step with the greatest increase in the mean RMS value. To find these points, the difference between the mean RMS at each time point and the time point 0.5 seconds later was computed as the change in mean RMS. Then, the MATLAB function \textit{findpeaks} was used to identify the time points with the greatest rate of change in muscle activation. These points were double-checked via visual inspection to avoid issues with any artifacts in the sEMG data. An example of this process can be seen in Fig. \ref{experimentoverview}(C).

For each subject and task type, 100 trial onset times were identified. The periods one second before these onset times were considered the `planning' stage when users would react to the stimuli and decide where to move. The periods one second after each trial onset were considered the `movement' stage when the user moved the robot handle to the desired position.

\subsubsection{Power Spectral Density Calculations}

After the EEG data was processed and segmented, the Power Spectral Density (PSD) was computed for each channel and time period. The PSD was computed via Welch's method, with a window length of 0.51 seconds and an overlap of 50\%.
For each channel, task, and user, the PSD was averaged across the 100 movement and planning periods to give the mean power for each user and channel during each of the two tasks directly before and after the movement onset.

The analysis of these results considered both the frequency band and brain region. The frequency band analysis focused on the delta (1-4 Hertz), theta (4-8 Hertz), and alpha (8-13 Hertz) bands.
For each channel, the relative power of the three frequency bands of interest was computed by summing the PSD of all frequencies in the band and dividing by the sum of the PSD at all frequencies in the range of 1 to 100 Hertz. 

Additionally, the relative power of the theta band compared to the alpha band was also considered. This was similarly computed by summing the PSDs of all frequencies in the theta band and dividing by the sum of the PSDs of all frequencies in the alpha band. This is referred to as the Theta-Alpha Ratio (TAR).

To compare how the PSD varied based on the area of the brain, the brain was segmented into six regions: frontal, central, parietal, occipital, and right and left temporal. The specific channels included in each region are shown in Fig. \ref{experimentoverview}(B). The mean relative PSDs from the channels in each region were taken for each subject and frequency band. 

In addition, the effect of the task duration on engagement was also investigated. For each subject, the data was split into thirds and the TARs were calculated separately for the first third and last third of trials. These results were compared for each task type and phase.

\subsubsection{Significance Testing}

The differences in PSD distributions for the Matching and Motor-Only tasks were considered during both the planning and movement phases. For each task, brain region, frequency band, and phase, the Kolmogorov-Smirnov test was performed over the distribution (across subjects) of relative PSD, and it failed to reject the null hypothesis that the data is normally distributed at a significance level of $p < 0.05$. Thus, the paired t-test was used to assess statistical significance between the relative frequencies. Four significance levels were considered for the paired t-test: 0.05, 0.01, 0.001, and 0.0001.

For the TARs, the Kolmogorov-Smirnov rejected the normality hypothesis for $p<0.05$, so the Wilcoxon signed-rank test was used to assess significance.

\section{Results and Discussion}

\subsection{Planning Phase}

\begin{figure}[h]
\centerline{\includegraphics[width=0.5\textwidth]{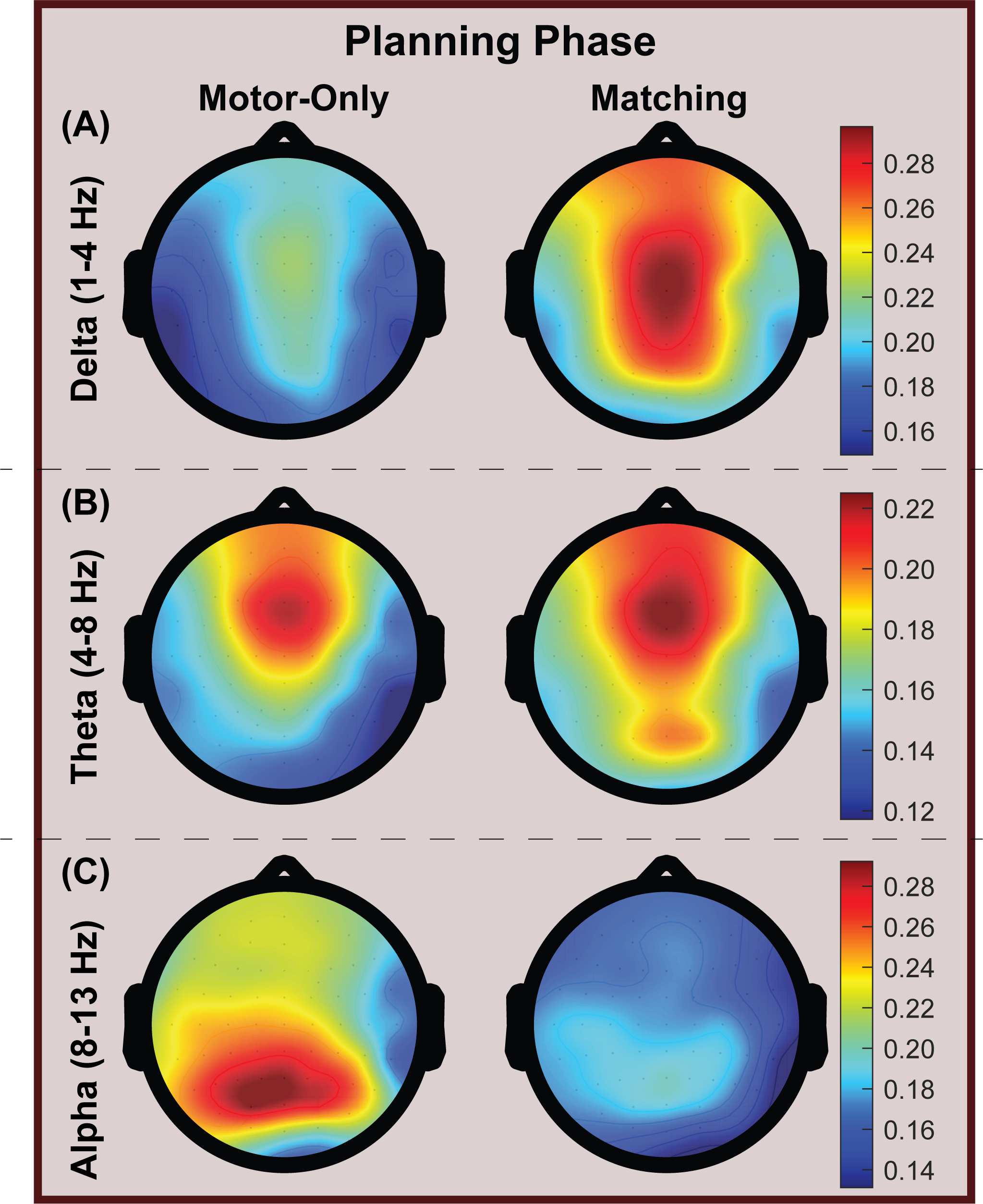}}
\caption{Mean relative PSD heatmaps of the brain during the planning phase of the Motor-Only and Matching tasks. Results are shown for (A) delta band, (B) theta band, and (C) alpha band.}
\label{BrainPlan}
\end{figure}

\begin{figure}[h]
\centerline{\includegraphics[width=0.5\textwidth]{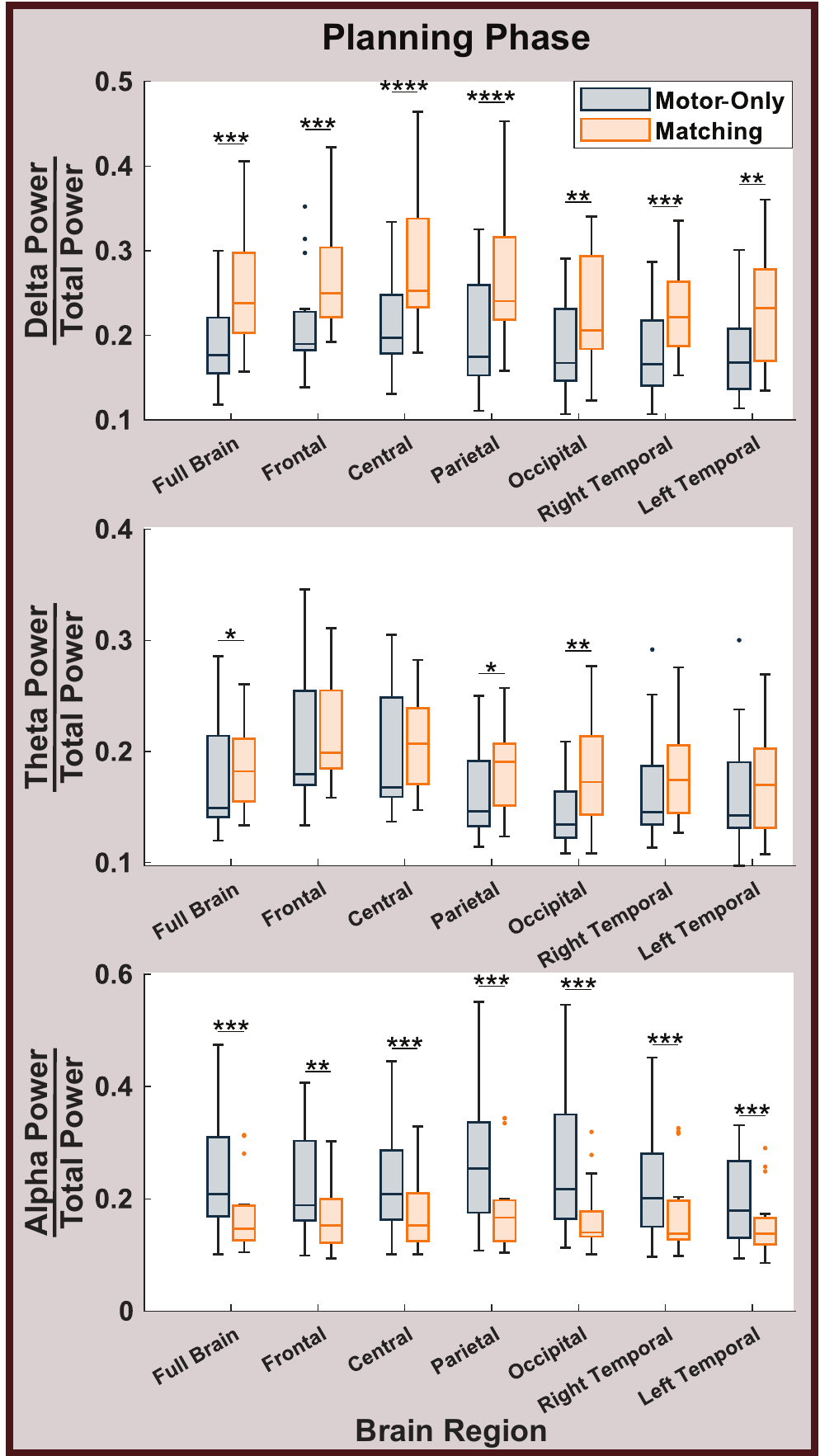}}
\caption{Boxplots showing relative PSD for different regions of the brain during the planning phase of the Motor-Only and Matching tasks. Results are shown for (A) delta band, (B) theta band, and (C) alpha band. Significant differences, per the paired t-test, are indicated with asterisks. *: $p<0.05$, **: $p<0.01$, ***: $ p<0.001$, and ****: $p<0.0001$.}
\label{BoxPlan}
\end{figure}

Heatmaps of the mean (across subjects) relative PSD of each of the three frequency bands during the planning phase are shown in Fig. \ref{BrainPlan}. The Motor-Only results are shown on the right, and the Matching results are on the right. 

Fig. \ref{BrainPlan}(A) shows the relative delta power of the Motor-Only and Matching tasks during the planning phase. The relative delta activity is much higher during the Matching task than during the Motor-Only task. In both tasks, the relative PSD is highest in the center of the head, centered around electrode Cz, which corresponds to activation in the motor cortex \cite{sibilla2018functional}. We can also see that the activity is higher in the frontal region during the Matching task, which corresponds to cognitive processing and attention \cite{sibilla2018functional}. Overall, the higher relative PSD in the delta region is indicative of greater attention and engagement during the Matching task.

In the theta band, shown in Fig. \ref{BrainPlan}(B), the change in relative PSD is not as great, with similar values in the frontal and central regions for both tasks. However, in the occipital region, there is a notable increase in relative theta PSD for the Matching task. This is particularly relevant because increased theta PSD in the occipital region has been specifically linked to engagement  \cite{davidesco2023detecting, lim2019comparison, magosso2019eeg}.

Considering the alpha band (Fig. \ref{BrainPlan}(C)), there is once again a clear difference between the relative PSDs of the Motor-Only and Matching tasks. However, unlike in the delta and theta bands, we find the relative alpha power is higher during the Motor-Only task than the Matching task. This decrease is particularly clear in the parietal and occipital regions, where relative PSD is very high in the Motor-Only task but much lower in the Matching task. For instance, the mean relative PSD at electrode POz is 0.28 for the Motor-Only task but only 0.19 for the Matching task. These observations are also indicative of an increase in engagement during the Matching task, as a decrease in alpha activity (particularly in the occipital and parietal regions) is strongly linked to increased engagement  \cite{davidesco2023detecting, lim2019comparison, magosso2019eeg}.

The occipital lobe is responsible for visual processing, and the parietal lobe is involved in sensation and perception \cite{sibilla2018functional}. As noted earlier, alpha activity is theorized to be an inhibitor for sensory processing \cite{magosso2019eeg, chikhi2022eeg}, so the decrease in occipital and parietal alpha power (and the corresponding increase in theta power) during the planning phase, when the user is paying attention to external stimuli and processing to see if the visual and auditory stimuli match, is expected in the Matching task. 

These observations from the heatmaps are supported quantitatively in Fig. \ref{BoxPlan}, which shows boxplots with the relative PSD distributions across subjects for each brain region and frequency band. For the delta and alpha bands, there is a significant difference in the relative PSD for all brain regions. As noted earlier, the PSD increases for the Matching task in the delta band and decreases for the Matching task in the alpha band. In both the alpha and delta bands, when considering the average PSD over the entire brain, 93\% of participants (14 out of 15) followed these trends. 

In the theta band, the difference between the PSD for the two tasks is significant only for the parietal and occipital regions and when averaging over the entire brain. When averaging over the entire brain, 80\% of the participants followed the trend of increased relative theta power during the Matching task compared to the Motor-Only task.

Altogether, the decreases in alpha power and increases in delta and theta power strongly support the hypothesis that the users are more engaged during the Matching task, as the trends in all three of the biomarkers show an increase in engagement.

\subsection{Movement Phase}

\begin{figure}[h]
\centerline{\includegraphics[width=0.5\textwidth]{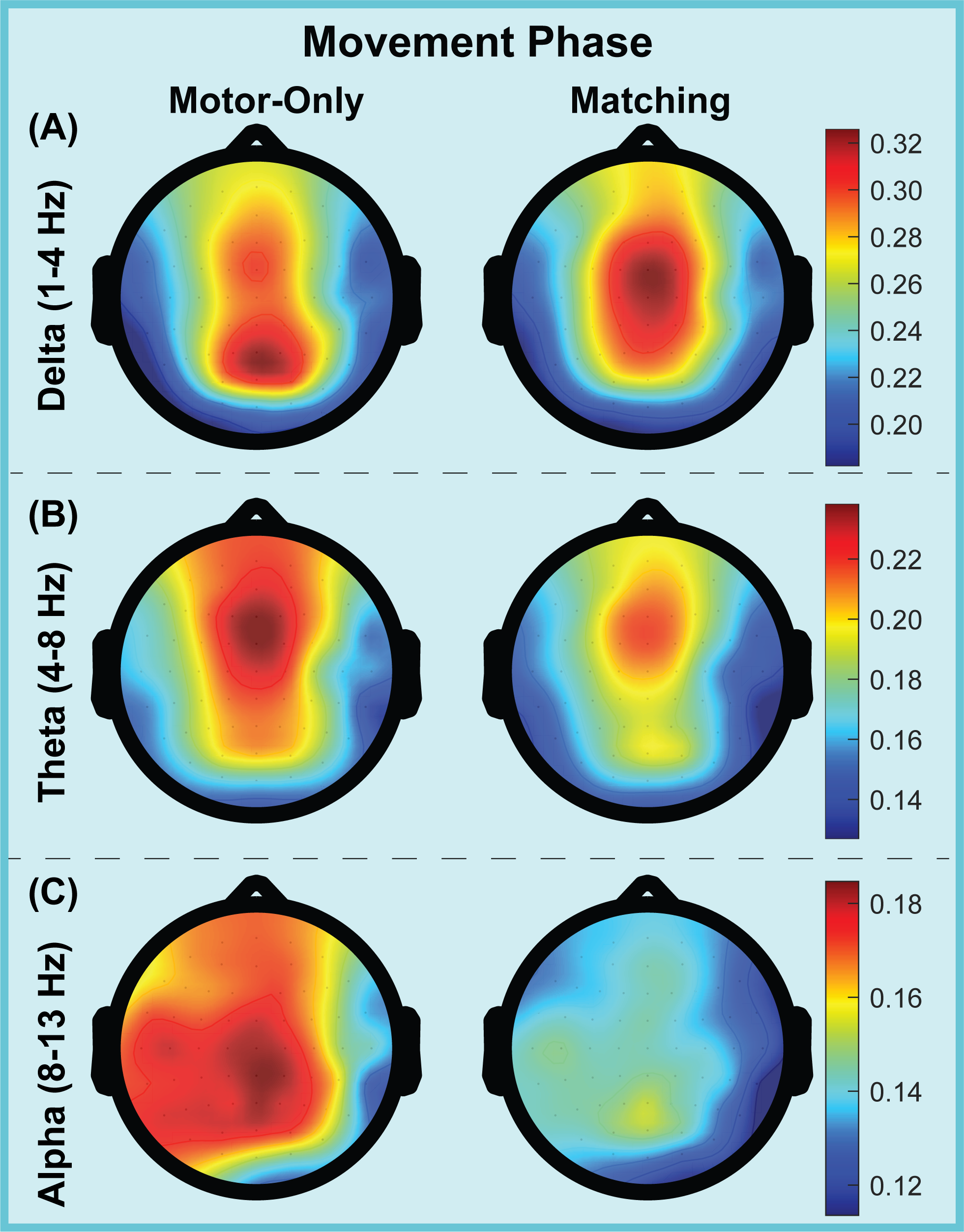}}
\caption{Mean relative PSD heatmaps of the brain during the movement phase of the Motor-Only and Matching tasks. Results are shown for (A) delta band, (B) theta band, and (C) alpha band.}
\label{BrainMove}
\end{figure}

\begin{figure}[h]
\centerline{\includegraphics[width=0.5\textwidth]{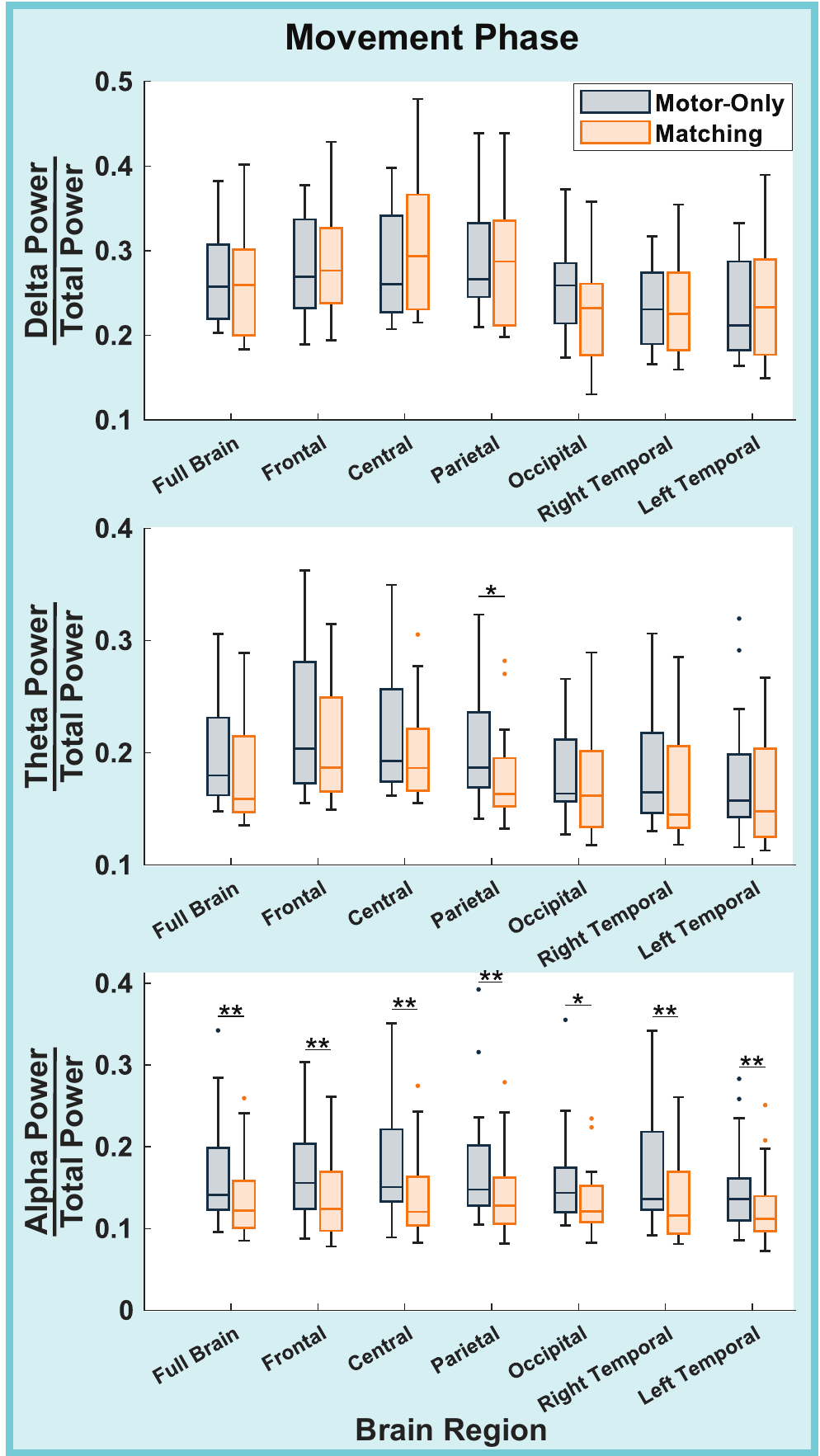}}
\caption{Boxplots showing relative PSD for different regions of the brain during the movement phase of the Motor-Only and Matching tasks. Results are shown for (A) delta band, (B) theta band, and (C) alpha band. Significant differences, per the paired t-test, are indicated with asterisks. *: $p<0.05$, **: $p<0.01$, ***: $ p<0.001$, and ****: $p<0.0001$.}
\vspace{-0.8cm}
\label{BoxMove}
\end{figure}

In the movement phase, the differences between the Motor-Only and Matching tasks are more subtle. The heatmaps and boxplots of these results are shown in Figs. \ref{BrainMove} and \ref{BoxMove}, respectively. 

The contrast between the planning and movement phases is most clear in the delta band. While there was a large difference between the two tasks in the planning phase, this is not true during the movement phase. Qualitatively considering Fig. \ref{BrainMove}(A), the relative delta power appears higher in the central region for the Matching task and in the occipital region for the Motor-Only task. However, these differences are not significant, and there are no significant differences in the delta power for any of the brain regions considered.

The trend has also changed in the theta band. While the theta power was higher for the Matching task in the planning phase, this is not the case for the movement phase. In fact, in this phase, the relative theta power is actually slightly higher in the Motor-Only task, though this difference only rises to the level of significance in the parietal region.

In contrast to the delta and theta bands, the relative alpha power follows the same trend for both the planning and movement phases. In both cases, the Matching task has significantly lower relative alpha power than the Motor-Only task for all regions of the brain. It should be noted that although the theta and delta bands are correlated with engagement, the alpha band activity is considered a more consistent indicator of engagement \cite{jin2019predicting, davidesco2023detecting}.

The similarities between the two tasks in delta and theta power during the movement phase may be explained by the similarities in the two tasks at this stage. In the planning phase, there is a substantial difference between the tasks, as the Matching task requires the user to process visual and auditory information and decide which direction to move, while in the Motor-Only task, the user only needs to see the target position. However, in the movement phase of both tasks, the user is completing the same action of moving the robot handle to the target location. Thus, since most of the sensory and cognitive processing is already complete, and the user is focused on motor control in this phase, the brain activity is similar between the two tasks.

\color{subsectioncolor}
\textcolor{subsectioncolor}{\subsection{Theta-Alpha Ratio}}
\color{black}

\begin{figure*}[ht]
\centerline{\includegraphics[width=0.88\textwidth]{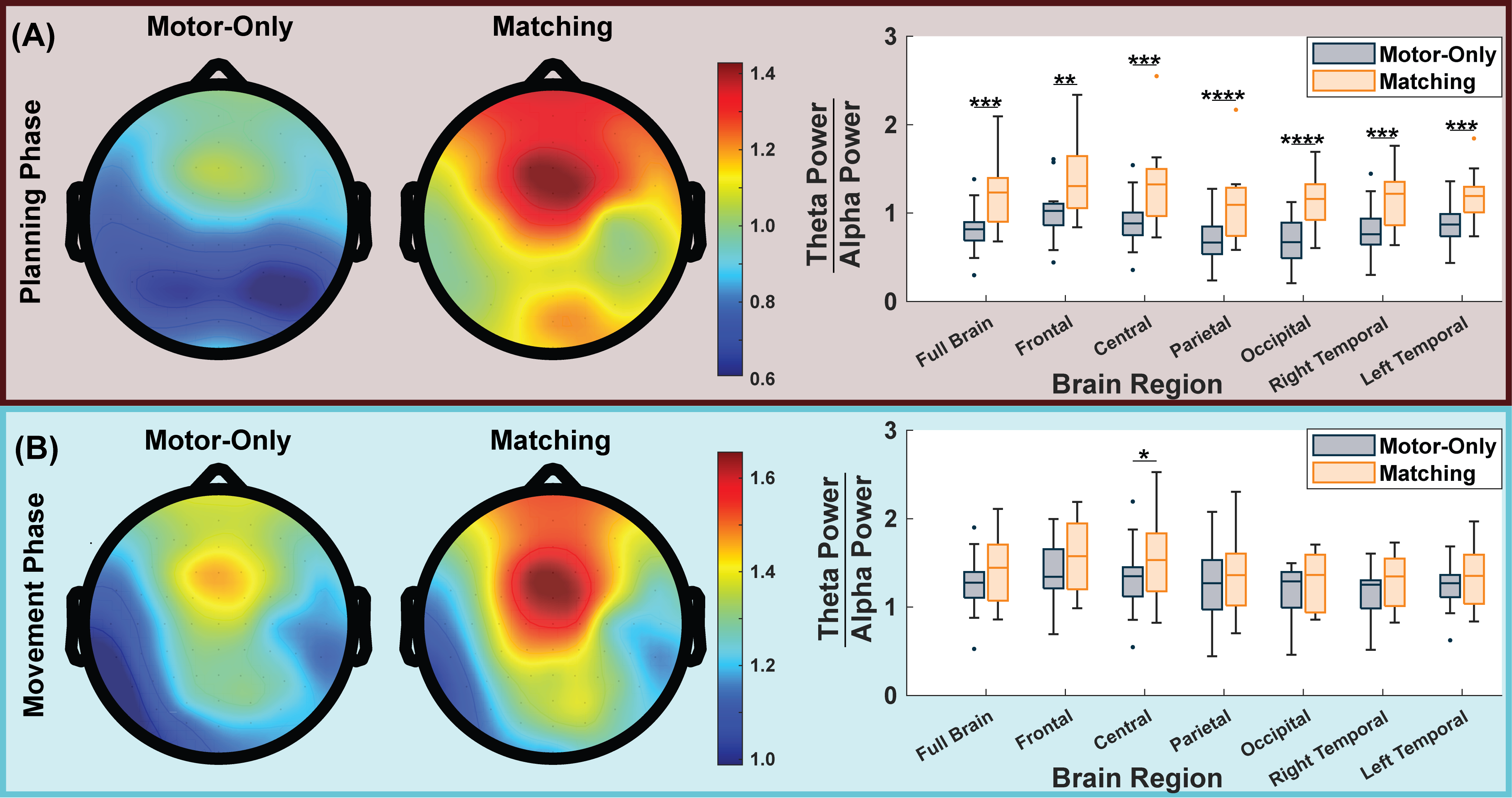}}
\caption{Boxplots and mean heatmaps for Theta-Alpha Ratios of the Motor-Only and Matching task for (A) the planning phase and (B) the movement phase. In the boxplots, significant differences, per the Wilcoxon signed-rank test, are indicated with asterisks. *: $p<0.05$, **: $p<0.01$, ***: $ p<0.001$, and ****: $p<0.0001$.}
\label{TAR}
\end{figure*}

Since theta power tends to increase with engagement, and alpha power tends to decrease with engagement, the ratio of theta power to alpha power can be used as a single metric that is sensitive to changes in both frequency bands \cite{lim2019comparison, nan2022alpha, pavithran2019index}. Fig. \ref{TAR} shows the mean Theta-Alpha Ratio (TAR) across the brain and the boxplots comparing the TAR distributions across subjects for different regions of the brain.

The difference between the Matching and Motor-Only tasks is evident in the heatmaps, which show higher TAR during the Matching task. This trend holds for both the planning and movement phases but is stronger during the planning phase. In the planning phase, the difference is statistically significant in every considered brain region, with 100\% of the subjects following the trend in the parietal and occipital regions and 93\% following the trend when considering all electrodes.

In the movement phase, the difference in TAR between the two tasks only achieves statistical significance for the central region. In the heatmaps, it can be seen that there is a clear increase in the TAR in this region, which corresponds to the motor cortex, responsible for movement \cite{sibilla2018functional}.
As stated above, the tasks are more similar to each other in the movement phase, so it follows that the brain activity would be more similar. The significant change in TAR during the planning phase indicates increased engagement during the Matching task.

\subsection{Changes to Engagement over Time}

\begin{figure}[ht!]
\centerline{\includegraphics[width=0.37\textwidth]{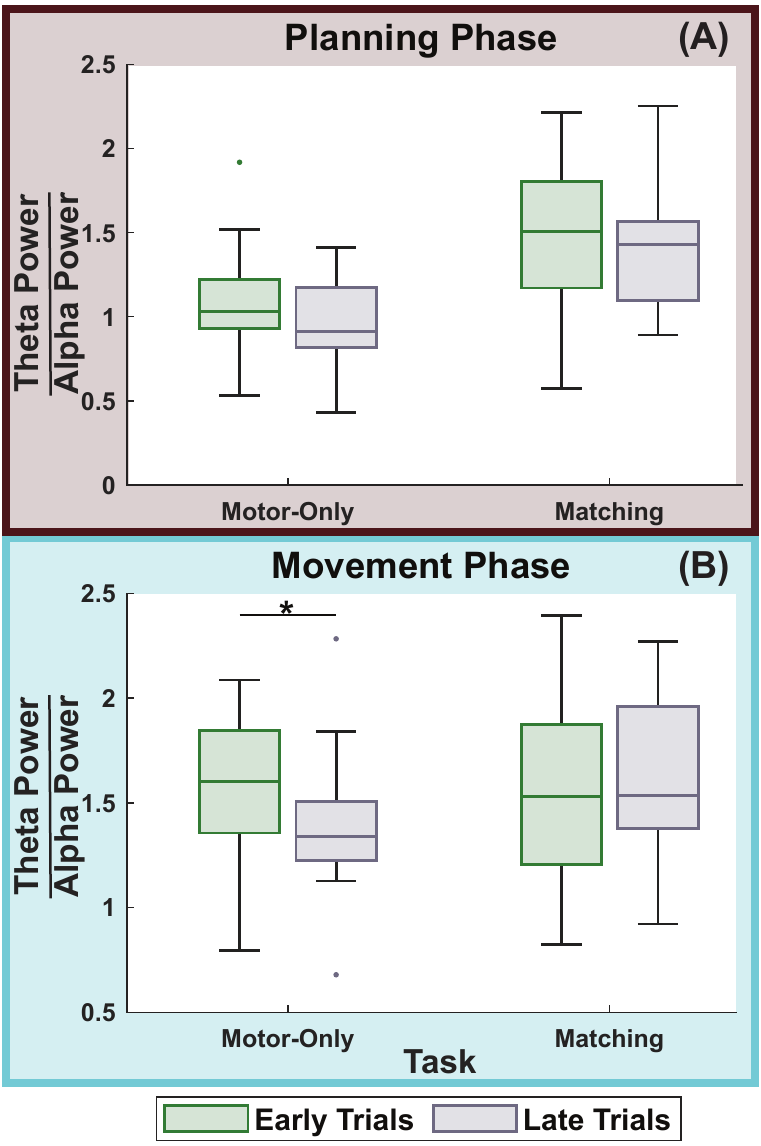}}
\caption{Boxplots comparing the Theta-Alpha Ratios for the first third of trials and the last third of trials for both the Motor-Only and Matching tasks. The results are shown for (A) the planning phase and (B) the movement phase. Significant differences ($p<0.05$) are indicating with an *.}
\label{TARChange}
\end{figure}

Another important aspect to consider is the ability to maintain user engagement over the course of the session. This is especially important in a rehabilitation context, where hundreds of repetitions can be needed to achieve the desired effect \cite{van2012systematic}, but users are likely to get bored and disengaged as a long session drags on. To investigate the impact of session time on this proposed motor learning regimen, we considered how the TAR changes for the users over the course of the session.

To this end, the TARs were computed using only the first third and last third of trials and compared. These results are shown in Fig. \ref{TARChange}. Since high TAR is indicative of high engagement, a drop in the TAR over the course of the session indicates that the user's engagement level is decreasing. 

In the Motor-Only task, the TAR tends to decrease as the task continues, with median changes of -11.5\% and -16.5\% for the planning and movement phases, respectively. However, this trend is only statistically significant in the movement phase (for $p<0.05$). This trend indicates that user engagement is decreasing as subjects do the trials, so a prolonged session of Motor-Only trials may yield diminishing returns as the user becomes disengaged.

In contrast, during the Matching task, the TARs are fairly consistent throughout the session, with a median change of -5.2\% and +0.4\% for the planning and movement phases, respectively. These differences are not significant in either phase. The consistent TAR levels indicate that the user's engagement level remains steady throughout the task, meaning they have a larger window for increased neuroplasticity and improved rehabilitation results.

\section{Conclusion}
User engagement increases neuroplasticity and participation in therapy sessions, making it a critical component of motor learning. Fifteen subjects took part in this study to investigate the effect of a multi-modal Matching task on user engagement. This new regimen was compared to a Motor-Only task, with participants completing one hundred repetitions of each trial type. EEG biomarkers for engagement, including relative power in the alpha, theta, and delta bands, as well as the ratio of theta power to alpha power, were considered. 
These biomarkers showed increased engagement during the new Matching task compared to the traditional Motor-Only task. These differences were particularly clear during the planning phase of each task when users were required to process visual and auditory stimuli in the Matching task. Additionally, while there was a decline in engagement over the course of the Motor-Only task, user engagement during the Matching task was consistent.
Engagement during motor learning is essential for patient improvement, as it increases neuroplasticity and motivates users to complete their therapy regimen. This work shows, for the first time, that a comprehensive robotic motor learning task involving the visual, auditory, and motor functions of the brain improves objective neural markers of engagement in healthy subjects. In future work, this proposed multi-modal motor learning approach should be evaluated on patients in need of rehabilitative therapy, with the ultimate goal of increasing the effectiveness of robotic rehabilitation.

\bibliographystyle{ieeetr}
\bibliography{citations}

\end{document}